\documentclass[amsmath,amssymb]{revtex4}
\usepackage{graphicx}
\usepackage{dcolumn}
\usepackage{bm}

\begin{document}

\preprint{APS/123-QED}

\title[Accounting for the large radial tension
   in wormholes]
{Accounting for the large radial tension in
   Morris-Thorne wormholes}
\author{Peter K.F. Kuhfittig}
\affiliation{Department of Mathematics\\
Milwaukee School of Engineering\\
Milwaukee, Wisconsin 53202-3109, USA}

\begin{abstract} It is well known that a
Morris-Thorne wormhole can only be held open
by violating the null energy condition,
physically realizable by the use of ``exotic
matter."  Unfortunately, even a small or
moderately-sized wormhole would have a radial
tension equal to that of the interior of a
massive neutron star. So outside a
neutron-star setting, such an
outcome is problematical at best, calling
for more than an appeal to exotic matter
whose introduction had a completely different
objective and with possibly different outcomes.
The purpose of this paper is to account for
the enormous radial tension in three ways:
(1) directly invoking noncommutative
geometry, an offshoot of string theory,
 (2) appealing to noncommutative geometry
in conjunction with $f(R)$ modified gravity,
and (3) determining the possible effect of
a small extra spatial dimension.
\end{abstract}

\maketitle

\vspace{12pt}

\section{Introduction}\label{S:Introduction}

Wormholes are handles or tunnels in spacetime
connecting widely separated regions of our
Universe or even different universes.  In
many ways, wormholes are as good a prediction
of Einstein's theory as black holes, but they
are subject to some severe restrictions.

Morris and Thorne \cite{MT88} proposed the
following static and spherically symmetric
line element for a wormhole spacetime:
\begin{equation}\label{E:line1}
ds^{2}=-e^{2\Phi(r)}dt^{2}+\frac{dr^2}{1-b(r)/r}
+r^{2}(d\theta^{2}+\text{sin}^{2}\theta\,
d\phi^{2}),
\end{equation}
using units in which $c=G=1$.
Here $\Phi =\Phi(r)$ is called the
\emph{redshift function}, which must be
everywhere finite to prevent an event
horizon.  The function $b=b(r)$ is called the
\emph{shape function} since it determines the
spatial shape of the wormhole when viewed,
for example, in an embedding diagram.  For
the shape function we must have $b(r_0)=r_0$,
where $r=r_0$ is the radius of the
\emph{throat} of the wormhole, and
$\text{lim}_{r\rightarrow\infty}b(r)/r=0$.
Another important requirement is the
\emph{flare-out condition} at the throat:
$b'(r_0)<1$, while $b(r)<r$ near the throat.
The flare-out condition can only be met by
violating the null energy condition (NEC)
\begin{equation}
   T_{\alpha\beta}k^{\alpha}k^{\beta}\ge 0
\end{equation}
for all null vectors $k^{\alpha}$, where
$T_{\alpha\beta}$ is the stress-energy tensor.
Matter that violates this condition is referred to
as ``exotic" in Ref. \cite{MT88}.  In particular,
for the radial outgoing null vector $(1,1,0,0)$,
the violation reads
$T_{\alpha\beta}k^{\alpha}k^{\beta}=\rho+
p_r<0$.  Here $T^t_{\phantom{tt}t}=-\rho(r)$
is the energy density, $T^r_{\phantom{rr}r}=
p_r(r)$ the radial pressure, and
$T^\theta_{\phantom{\theta\theta}\theta}=
T^\phi_{\phantom{\phi\phi}\phi}=p_t(r)$
the lateral pressure.

Next, let us state the Einstein field
equations:
\begin{equation}\label{E:Einstein1}
  \rho(r)=\frac{b'}{8\pi r^2},
\end{equation}
\begin{equation}\label{E:Einstein2}
   p_r(r)=\frac{1}{8\pi}\left[-\frac{b}{r^3}+
   2\left(1-\frac{b}{r}\right)\frac{\Phi'}{r}
   \right],
\end{equation}
and
\begin{equation}\label{E:Einstein3}
   p_t(r)=\frac{1}{8\pi}\left(1-\frac{b}{r}
   \right)
   \left[\Phi''-\frac{b'r-b}{2r(r-b)}\Phi'
   +(\Phi')^2+\frac{\Phi'}{r}-
   \frac{b'r-b}{2r^2(r-b)}\right].
\end{equation}
We also need to recall that the radial
tension $\tau(r)$ is the negative of the
radial pressure $p_r(r)$.

Having established the need for exotic
matter, Morris and Thorne go on to state
that the Einstein field equations can be
rearranged to yield $\tau$: reintroducing
$c$ and $G$ for now, this is given by
\begin{equation}\label{E:tau}
   \tau(r)=\frac{b(r)/r-2[r-b(r)]\Phi'(r)}
   {8\pi Gc^{-4}r^2}.
\end{equation}
From this condition it follows that the
radial tension at the throat is
\begin{equation}\label{E:tension1}
   \tau=\frac{1}{8\pi Gc^{-4}r_0^2}\approx
   5\times 10^{41}\frac{\text{dyn}}{\text{cm}^2}
   \left(\frac{10\,\text{m}}{r_0}\right)^2.
\end{equation}
In particular, for $r_0=3\,\,\text{km}$,
$\tau$ has the same magnitude as the pressure
at the center of a massive neutron star. Such
a situation has already been considered.
According to Ref. \cite{pK12a}, the enormous
pressure near the center could cause a wormhole
to form, provided that there is a core of quark
matter.

To see how these ideas fit together, let us
return to the null vector $(1,1,0,0)$ and
observe that $\rho +p_r<0$ can be written
$\tau -\rho >0$.  In the nongeometrized
units of Eq. (\ref{E:tau}), the
inequality becomes $\tau >\rho c^2$,
possibly leading to an enormous radial
tension at the throat.  On the other hand,
if $\rho$ is extremely small, then $\tau$
can be quite small, as well.  An even
more revealing case, the zero-density
solution corresponding to a constant
shape function [see Eq.
(\ref{E:Einstein1})], is discussed by
Visser \cite{mV95}.  Similarly, if $r_0$
is extremely large, then, according to Eq.
(\ref{E:tau}), $\tau$ may again be small.
So while exotic matter ensures the
violation of the NEC, a large radial
tension is not a defining property.  In
particular, if we are not dealing with
the interior of a neutron star, then
the excessive radial tension becomes
hard to explain.  The usual, often
unstated, assumption is that the
extreme conditions are due to exotic
matter.  This assumption now takes on
the appearance of merely giving the
mystery a name.

The purpose of the paper is to account for
the enormous radial tension in three ways:
(1) directly invoking noncommutative
geometry, itself an offshoot of string
theory, (2) appealing to noncommutative
geometry in conjunction with $f(R)$
modified gravity, and (3) analyzing the
effect of a small extra spatial dimension.

\section{Noncommutative geometry}
   \label{S:noncommutative}
There is no doubt that string theory has
become ever more influential, as exemplified
by the realization that coordinates may
become noncommutative operators on a
$D$-brane \cite{eW96, SW99}.  The result,
noncommutative geometry, can help eliminate
the divergences that normally occur in
general relativity.  The reason is that
noncommutativity replaces point-like
objects by smeared objects, as a result
of which spacetime can be encoded in the
commutator
$[\textbf{x}^{\mu},\textbf{x}^{\nu}]
=i\theta^{\mu\nu}$, where $\theta^{\mu\nu}$ is
an antisymmetric matrix that determines the
fundamental cell discretization of spacetime
in the same way that Planck's constant $\hbar$
discretizes phase space \cite{NSS06}.

A natural way to model the smearing effect is
by means of a Gaussian distribution of minimal
length $\sqrt{\beta}$ instead of the Dirac
delta function \cite{NSS06, NS10, mR11, RKRIb,
pK13}.  An equally effective way, discussed in
Refs. \cite{LL12, NM08, pK12}, is to assume
that the energy density of the static and
spherically symmetric and particle-like
gravitational source has the form
\begin{equation}\label{E:rho1}
  \rho_{\beta}(r)=\frac{\mu_1\sqrt{\beta}}
     {\pi^2(r^2+\beta)^2},
\end{equation}
where $\mu_1$ is a constant.  Eq. (\ref{E:rho1})
can be interpreted to mean that the
gravitational source causes the mass $\mu_1$ of
a particle to be diffused throughout a
region of linear dimension $\sqrt{\beta}$
due to the uncertainty; so $\sqrt{\beta}$
has units of length.  Following Ref.
\cite{NSS06}, Eq. (\ref{E:rho1}) likewise
leads to the mass distribution
\begin{equation}
   m(r)=\int^r_04\pi(r')^2\rho(r')dr'
   =\frac{2M}{\pi}\left(\text{tan}^{-1}
   \frac{r}{\sqrt{\beta}}-
   \frac{r\sqrt{\beta}}{r^2+\beta}\right),
\end{equation}
where $M$ is now the total mass of the
source.  Observe that $m(0)=0$ but $m(r)$
rapidly rises to $M$ as $r$ increases.

According to Ref. \cite{NSS06}, noncommutative
geometry is an intrinsic property of spacetime
and does not depend on any particular features
such as curvature.  Moreover, the relationship
between the radial pressure and energy density
is given by
\begin{equation}\label{E:pressure}
  p_r=-\rho_{\beta}.
\end{equation}
The reason is that the source is a
self-gravitating droplet of anisotropic
fluid of density $\rho_{\beta}$ and the
radial pressure is needed to prevent a
collapse to a matter point.

In Ref. \cite{NSS06}, the line element
describing a black hole is obtained by
solving the Einstein field equations with
the Gaussian distribution as a matter
source.  Momentarily returning to Eq.
(\ref{E:rho1}), by using this form instead
of the Gaussian distribution, the line
element describing a black hole becomes
\begin{multline}\label{E:line4}
   ds^2=-\left[1-\frac{(4M/ \pi)\left(\text{tan}^{-1}
   \frac{r}{\sqrt{\beta}}-\frac{r\sqrt{\beta}}
   {r^2+\beta}\right)}{r}\right]dt^2\\
   +\left[1-\frac{(4M/ \pi)\left(\text{tan}^{-1}
   \frac{r}{\sqrt{\beta}}-\frac{r\sqrt{\beta}}
   {r^2+\beta}\right)}{r}\right]^{-1}dr^2
   +r^{2}(d\theta^{2}+\text{sin}^{2}\theta
    \,d\phi^{2}).
\end{multline}
As $\beta\rightarrow 0$, we recover the
Schwarzschild line element.

Regarding the event horizon, due to the
uncertainty, we are now dealing with a
smeared surface instead of a smeared
particle.  These ideas carry over to
wormholes in the sense that the throat also
becomes a smeared surface.

\section{Noncommutative-geometry inspired wormholes}
\label{S:inspired}
\subsection{The shape function}

Since we are now concerned with the effects
of noncommutative geometry, we have to
return to Ref. \cite{NSS06} to discuss the
length scales.  It is stated that, at first
glance, there appears to be a need to modify
the 4D Einstein action to incorporate the
noncommutative effects.  We are dealing
here with a very small scale which
is definitely needed to account for
the large radial tension at the throat of
any moderately-sized Morris-Thorne wormhole.
At the same time, however, we are dealing
with a geometric structure over an
underlying manifold which affects gravity
in a subtle and indirect way: noncommutativity
influences the mass-energy and momentum
distribution and propagation \cite{SS02,
SS03a, SS03b}, while the energy-momentum
density determines the spacetime curvature.
It is concluded in Ref. \cite{NSS06} that
in general relativity, the effects of
noncommutativity can be taken into account
by keeping the standard form of the
Einstein tensor on the left-hand side of
the field equations and introducing the
modified energy-momentum tensor as a source
on the right-hand side.  This leads to the
conclusion, also discussed in Ref.
\cite{RJ16}, that the length scales need not
be microscopic,  which, in turn, allows us
to obtain the shape function by means of
Eqs. (\ref{E:Einstein1}) and
(\ref{E:rho1}), but it does not prevent us
from examining the region near the throat.
The shape function is given by
\begin{equation}\label{E:shape}
  b(r)=\frac{8M\sqrt{\beta}}{\pi}\int^r_{r_0}
  \frac{(r')^2dr'}{[(r')^2+\beta]^2}+r_0
  =\frac{4M\sqrt{\beta}}{\pi}
  \left(\frac{1}{\sqrt{\beta}}\text{tan}^{-1}
  \frac{r}{\sqrt{\beta}}-\frac{r}{r^2+\beta}
  -\frac{1}{\sqrt{\beta}}\text{tan}^{-1}
  \frac{r_0}{\sqrt{\beta}}+\frac{r_0}{r_0^2
  +\beta}\right)+r_0.
\end{equation}
According to Ref. \cite{pK15}, $b=b(r)$ has
the usual properties of a shape function:
$b(r_0)=r_0$, $0<b'(r_0)<1$, and
$\text{lim}_{r\rightarrow\infty}b(r)/r=0$.

Based on the above discussion, we are allowed
to use Eq. (\ref{E:rho1}) to obtain a wormhole
solution on a macroscopic scale.  The fact
remains, however, that $\rho_{\beta}(r)$ in
Eq. (\ref{E:rho1}) describes a particle-like
gravitational source near the origin.  To
study the energy density farther away from
the origin, some additional assumptions will
have to be made.  That is the topic of the
next subsection.

\subsection{The throat surface}
As noted in Sec. \ref{S:noncommutative}, we
assume that the throat $r=r_0$ is a smeared
surface.  The best way to describe such a
surface is to imagine that the smeared particle
at $r=0$ is replaced by a set of smeared
particles on the surface, which is precisely
what causes the surface to be smeared.  Now
let us consider the same outward radial
direction for the particle and surface, as
shown in Fig. 1.  The two labeled line
\begin{figure}[tbp]
\begin{center}
\includegraphics[width=0.5\textwidth]{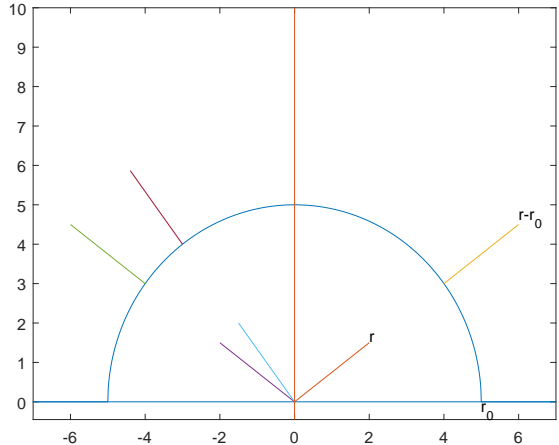}
\end{center}
\caption{Section of a sphere of radius $r=r_0$.}
\end{figure}
segments  were drawn with the same length and
direction to show the correspondence between
the particle at $r=0$ and the particle on the
surface.  As a consequence, it is reasonable
to assume that the energy density $\rho_s$
of the smeared surface is a simple translation
of Eq. (\ref{E:rho1}) in the $r$-direction:
\begin{equation}\label{E:rho2}
   \rho_s=\frac{\mu_2\sqrt{\beta}}{\pi^2
   [(r-r_0)^2+\beta]^2},
\end{equation}
where $\mu_2$ is the mass of the surface.
Moreover, according to Ref. \cite{NSS06},
we are dealing here with a kind of quantum
pressure, i.e., an outward push induced by
noncommuting coordinate quantum
fluctuations, which allow us to go from
point-particle to surface.  We also need
to recall that $r_0$ is a relatively
moderate throat radius, such as the 3-km
radius discussed in Sec. \ref
{S:Introduction}.  Finally, Eq.
(\ref{E:pressure}) carries over to the
surface, as long as we confine ourselves
to the outward radial direction.

Since the tension $\tau$ is the negative of
$p_r$, the violation of the NEC, $p_r+\rho_s
<0$, now becomes $ \tau -\rho_s>0$ at the
throat.  The condition $p_r=-\rho_s$
implies that we are right on the edge of
violating the NEC, i.e., $\tau -\rho_s=0$.
In particular, when $r=r_0$,
\begin{equation}\label{E:rhos}
  \rho_s=\frac{\mu_2}{\pi^2}\frac{1}{\beta^{3/2}},
\end{equation}
but we still have $\tau -\rho_s=0$.  However,
since the throat is a smeared surface, we
can only assert that $r\approx r_0$.  So by
Eq. (\ref{E:rho2}), $\rho_s$ is reduced in
value and we obtain the desired
\begin{equation}
  \tau -\rho_s>0.
\end{equation}
Finally, Eq. (\ref{E:rhos}) implies that
$\rho_s$ and hence $\tau$ are extremely
large near the throat.  This is our main
result, to be confirmed later by other
means.

To check its plausibility, let us return
to Sec. \ref{S:Introduction}, Eq.
(\ref{E:tension1}), and recall that for
a throat size of 10 m,
$\tau\approx 5\times 10^{41} \,\text{dyn}
/\text{cm}^2$. Suppose $\mu_2$ has the
rather minute value $10^{-10}$ g.  Applying
Eq. (\ref{E:rhos}), we have
\begin{equation}
  \tau=\rho_sc^2=\frac{\mu_2}{\pi^2}
  (\sqrt{\beta})^{-3}c^2=5\times 10^{41}
  \,\frac{\text{dyn}}{\text{cm}^2}.
\end{equation}
To satisfy this relationship, the value
$\sqrt{\beta}=10^{-11}\,\text{cm}$ is
sufficient.  Since $\sqrt{\beta}$ may be
much smaller, we can accommodate even
larger values of $\tau$.

\section{Connection to $f(R)$ modified
gravity}

Modified gravitational theories, including
$f(R)$ modified gravity, have been invoked
to explain various phenomena in general
relativity.  In this section we present a
heuristic argument to show that $f(R)$
modified gravity cannot provide a
satisfactory explanation for the extreme
radial tension at the throat without
appealing to noncommutative geometry.

First we need to recall that in $f(R)$
gravity, the Ricci scalar $R$ in the
Einstein-Hilbert action $S_{\text{EH}}=
\int\sqrt{-g}\,R\,dx^4$ is replaced by
a nonlinear function $f(R)$.  Next,
let us state the gravitational field
equations in the form used by Lobo and
Oliveira \cite{LO09}:
\begin{equation}\label{E:Lobo1}
   \rho(r)=F(r)\frac{b'(r)}{r^2},
\end{equation}
\begin{equation}
   p_r(r)=-F(r)\frac{b(r)}{r^3}+F'(r)
   \frac{rb'(r)-b(r)}{2r^2}-F''(r)
   \left[1-\frac{b(r)}{r}\right],
\end{equation}
and
\begin{equation}
   p_t(r)=-\frac{F'(r)}{r}\left[1-
   \frac{b(r)}{r}\right]+\frac{F(r)}
   {2r^3}[b(r)-rb'(r)];
\end{equation}
here $F=\frac{df}{dR}$.  It is also
assumed that $\Phi(r)\equiv\text{constant}$,
so that $\Phi'(r)\equiv 0$.  Otherwise,
according to Ref. \cite{LO09}, the
analysis becomes intractable.  Fortunately,
the redshift function does not have to be
specified in our discussion.

Returning to the field equations, at
$r=r_0$, we have
\begin{equation}
   \rho(r_0)=F(r_0)\frac{b'(r_0)}{r_0^2}
\end{equation}
and
\begin{equation}
   \tau(r_0)=-p_r(r_0)=F(r_0)\frac{b(r_0)}
   {r_0^3}-F'(r_0)\frac{r_0b'(r_0)-b(r_0)}
   {2r_0^2}.
\end{equation}
To produce a large radial tension, it is
sufficient for $F'(r_0)$ to be large and
positive.  $F(r_0)$ itself can be quite
small and still meet the flare-out
condition $b'(r_0)=r_0^2\rho(r_0)/F(r_0)
<1$.  The result is a large tension, while
retaining the small $\rho(r_0)$.

The real question is whether the positive
conclusion from $f(R)$ gravity can be
considered satisfactory.  While $F(r_0)$
does not need to meet any special
requirements, the very large $F'(r_0)$
demanded seems rather artificial, even
unphysical.  Fortunately, we can still
fall back on our previous
noncommutative-geometry  theory.

To that end, we need to refer to Ref.
\cite{pK18}, which discusses the connection
between noncommutative geometry and $f(R)$
modified gravity.  According to Ref. \cite
{LO09}, the Ricci scalar $R$ is given by
$R=2b'(r)/r$.  So in the vicinity of the
throat, $R\approx 2\alpha/r^2$, for
$\alpha <1$. To be consistent with the
notation in Ref. \cite {pK18}, we will
assume the following form for the Ricci
scalar:
\begin{equation}\label{E:R}
   R(r)=\frac{2\alpha}{r^2},
\end{equation}
where $\alpha$ is a constant.  It is shown
in Ref. \cite{pK18} that the
noncommutative-geometry background yields
the form
\begin{equation}
   f(R)=\frac{2M_{r_0}}{\pi^2\beta^{3/2}}
   \,\text{ln}\left(1+\frac{\beta R}{2\alpha}
   \right),
\end{equation}
where $M_{r_0}$ is the mass of a
spherical shell of radius $r=r_0$, which
we can take to be  the surface of the
throat.  Given that
$\beta R/2\alpha=\beta/r^2<1$, the Maclaurin
expansion now yields
\begin{equation}
  f(R)\approx\frac{2M_{r_0}}
  {\pi^2\beta^{3/2}}\left[\frac{\beta R}{2\alpha}
  -\frac{1}{2}\left(\frac{\beta R}{2\alpha}
  \right)^2\right]<\frac{2M_{r_0}}
  {\pi^2\beta^{3/2}}\frac{\beta R}{2\alpha}
  =\frac{M_{r_0}}{\pi^2\alpha\sqrt{\beta}}R.
\end{equation}
So
\begin{equation}\label{E:agree}
   f'(R)<\frac{M_{r_0}}{\pi^2\alpha\sqrt{\beta}}.
\end{equation}
This result agrees with the more convenient
assumption that $f(R)$ has the approximate
form
\begin{equation}
   f(R)=\frac{M_{r_0}}{\pi^2\alpha\sqrt{\beta}}
   R^{1-\varepsilon},\quad \varepsilon
   \ll 1.
\end{equation}
To show this, observe that for $\varepsilon
=0$, $(1-\varepsilon)R^{-\varepsilon}=1$,
while $\frac{\partial}{\partial\epsilon}
(1-\varepsilon)R^{-\varepsilon}<0$.  So
for $\varepsilon >0$, $(1-\varepsilon)
R^{-\varepsilon}<1$.  We now have
\begin{equation}
   f'(R)=\frac{M_{r_0}}{\pi^2\alpha
   \sqrt{\beta}}
   (1-\varepsilon)R^{-\varepsilon}<
   \frac{M_{r_0}}{\pi^2\alpha\sqrt{\beta}},
\end{equation}
in agreement with Inequality
(\ref{E:agree}).

Combining these observations, we have
\begin{equation*}
   F(R)=F=\frac{df}{dR}=
   \frac{M_{r_0}}{\pi^2\alpha\sqrt{\beta}}
   (1-\varepsilon)R^{-\varepsilon},
\end{equation*}
\begin{equation*}
   \frac{dF}{dR}=
   \frac{M_{r_0}}{\pi^2\alpha\sqrt{\beta}}
   (1-\varepsilon)(-\varepsilon)
   R^{-\varepsilon-1},
\end{equation*}
and
\begin{equation*}
   \frac{dR}{dr}=-\frac{4\alpha}{r^3}.
\end{equation*}
The result is
\begin{equation*}
   \frac{dF}{dr}=\frac{dF}{dR}
   \frac{dR}{dr}=
   \frac{M_{r_0}}{\pi^2\alpha\sqrt{\beta}}
   (1-\varepsilon)(-\varepsilon)
   R^{-\varepsilon-1}\left(-\frac{4\alpha}{r^3}
   \right)>0
\end{equation*}
near the throat $r=r_0$.  So $dF/dr$ is
large as long as $\sqrt{\beta}\ll M_{r_0}$.
We conclude that $F'(r_0)$ can indeed be
large and positive due to the
noncommutative-geometry background.

The connection between noncommutative
geometry and $f(R)$ modified gravity
can be viewed from a broader perspective.
From Eq. (\ref{E:Lobo1}) with $b'(r)=
2\alpha$, \begin{equation}\label{E:mod1}
   F(r)=\frac{r^2}{b'(r)}\rho(r)=
   \frac{r^2}{2\alpha}\rho(r)=
   \frac{1}{2\alpha/r^2}
   \frac{\mu_1\sqrt{\beta}}
   {\pi^2(r^2+\beta)^2},
\end{equation}
while $r(R)=\sqrt{2\alpha/R}$ from
Eq. (\ref{E:R}).  Substituting in Eq.
(\ref{E:mod1}), we get
\begin{equation}\label{E:mod2}
   F(R)=\frac{\mu_1\sqrt{\beta}}
   {\pi^2}\frac{1}
   {R\left(\frac{2\alpha}{R}+\beta
   \right)^2}.
\end{equation}
Since $F=\frac{df}{dR}$, Eq.
(\ref{E:mod2}) yields
\begin{equation}\label{E:mod3}
    f(R)=\int^R_0\frac{\mu_1\sqrt{\beta}}{\pi^2}
    \frac{1}{R'\left(\frac{2\alpha}{R'}+\beta
    \right)^2}dR'=
    \frac{\mu_1\sqrt{\beta}}{\pi^2}
    \frac{(\beta R+2\alpha)\text{ln}\,
    (\beta R+2\alpha)-\beta R}
    {\beta^2(\beta R+2\alpha)}+C.
\end{equation}
Lobo and Oliveira \cite{LO09} go on to
show that the matter threading the
wormhole satisfies the null energy
condition, while the effective
energy-momentum tensor arising from the
modified gravity theory leads to a
violation, thereby supporting the
wormhole.

The result is of interest to us
because Eq. (\ref{E:mod3}) is obtained
from noncommutative geometry and
therefore provides a motivation for
the actual choice of the function
$f(R)$.
\\
\\
\section{The extra spatial dimension}

In discussing an extra spatial dimension,
we are going to follow Ref. \cite
{pK18a}, which starts with a generic
line element in Schwarzschild
coordinates:
\begin{equation}\label{E:L1}
  ds^2=-e^{2\Phi(r)}dt^2
  +e^{2\lambda(r)}dr^2 +r^2(d\theta^2
  +\text{sin}^2\theta\,d\phi^2).
\end{equation}
Here it is assumed that from symmetry
considerations, the fifth component has
the form $e^{2\mu(r,l)}dl^2$, where $l$
is the extra coordinate.  In a wormhole
setting, we naturally assume that
$e^{2\lambda(r)}=1-b(r)/r$, resulting
in the form
\begin{equation}\label{E:L4}
  ds^2=-e^{2\Phi(r)}dt^2+\frac{dr^2}{1-b(r)/r}+r^2
  (d\theta^2+\text{sin}^2\theta\,d\phi^2)
  +e^{2\mu(r,l)}dl^2,
\end{equation}
again assuming that $c=G=1.$

Ref. \cite{pK18a} is concerned primarily
with the null energy condition.  As already
noted, the purpose of this paper is to
account for the extreme radial tension at
the throat.

Since Ref. \cite{pK18a} uses an orthonormal
frame, we need to introduce the following
notations for the components of the
stress-energy tensor: $T_{00}=\rho$, the
energy density, $T_{11}=p_r$, the radial
pressure, and $\tau =-p_r$, the radial
tension.  Still following Ref. \cite
{pK18a}, we choose an orthonormal basis
$\{e_{\hat{\alpha}}\}$ which is dual to the
following 1-form basis:
\begin{equation}\label{E:oneform1}
    \theta^0=e^{\Phi(r)}\, dt,
    \quad \theta^1=[1-b(r)/r]^{-1/2}\,dr,
     \quad\theta^2=r\,d\theta, \quad
      \theta^3=r\,
  \,\text{sin}\,\theta\,d\phi,
  \quad \theta^4=e^{\mu(r,l)}dl.
\end{equation}
The resulting 1- and 2-forms and the
derivations of the Riemann curvature and
Ricci tensors given in Ref. \cite{pK18a}.
For convenience, let us restate the
components of the Ricci tensor in the
orthonormal frame:
\begin{multline}
  R_{00}=-\frac{1}{2}\frac{d\Phi(r)}{dr}\frac{rb'-b}{r^2}
  +\frac{d^2\Phi(r)}{dr^2}\left(1-\frac{b}{r}\right)\\+
  \left[\frac{d\Phi(r)}{dr}\right]^2\left(1-\frac{b}{r}\right)
  +\frac{2}{r}\frac{d\Phi(r)}{dr}\left(1-\frac{b}{r}\right)
  +\frac{d\Phi(r)}{dr}\frac{\partial\mu(r,l)}{\partial r}
  \left(1-\frac{b}{r}\right),
\end{multline}
\begin{multline}
   R_{11}=\frac{1}{2}\frac{d\Phi(r)}{dr}\frac{rb'-b}{r^2}
  -\frac{d^2\Phi(r)}{dr^2}\left(1-\frac{b}{r}\right)\\
  -\left[\frac{d\Phi(r)}{dr}\right]^2\left(1-\frac{b}{r}\right)
  +\frac{rb'-b}{r^3}-\frac{\partial^2\mu(r,l)}{\partial r^2}
  \left(1-\frac{b}{r}\right)\\
  +\frac{1}{2}\frac{\partial\mu(r,l)}{\partial r}
  \frac{rb'-b}{r^2}-\left[\frac{\partial\mu(r,l)}{\partial r}
  \right]^2\left(1-\frac{b}{r}\right),
\end{multline}
\begin{equation}
   R_{22}=R_{33}=-\frac{1}{r}\frac{d\Phi(r)}{dr}
   \left(1-\frac{b}{r}\right)+\frac{1}{2}\frac{rb'-b}{r^3}
   +\frac{b}{r^3}
   -\frac{1}{r}\frac{\partial\mu(r,l)}{\partial r}
   \left(1-\frac{b}{r}\right),
\end{equation}
\begin{multline}
  R_{44}=-\frac{d\Phi(r)}{dr}\frac{\partial\mu(r,l)}{\partial r}
  \left(1-\frac{b}{r}\right)-\frac{\partial^2\mu(r,l)}
  {\partial r^2}\left(1-\frac{b}{r}\right)\\
  +\frac{1}{2}\frac{\partial\mu(r,l)}{\partial r}\frac{rb'-b}{r^2}
  -\left[\frac{\partial\mu(r,l)}{\partial r}\right]^2
  \left(1-\frac{b}{r}\right)
  -\frac{2}{r}\frac{\partial\mu(r,l)}{\partial r}
  \left(1-\frac{b}{r}\right).
\end{multline}
Since we are primarily interested in the
radial tension and pressure at the throat,
where $b(r_0)=r_0$, we may assume that
$1-b/r=0$ in all these expressions.

To study the radial tension $\tau(r)=
-p_r(r)=-T_{11}$, we start with the
Einstein field equations in the
orthonormal frame:
\begin{equation}
   G_{\alpha\beta}
   =R_{\alpha\beta}-\frac{1}
{2}Rg_{\alpha\beta}
=8\pi T_{\alpha\beta};
\end{equation}
thus

\begin{equation}
   g_{\alpha\beta}=
   \left(
   \begin{matrix}
   -1&0&0&0&0\\
   \phantom{-}0&1&0&0&0\\
   \phantom{-}0&0&1&0&0\\
   \phantom{-}0&0&0&1&0\\
   \phantom{-}0&0&0&0&1
   \end{matrix}
   \right).
\end{equation}
From $G_{11}=8\pi T_{11}$ we obtain
\begin{equation}
   8\pi p_r(r)=R_{11}-\frac{1}{2}Rg_{11},
\end{equation}
where the Ricci scalar is given by
\begin{equation}\label{E:Ricci}
R=R^i_{\phantom{0}i}=-R_{00}+R_{11}
  +R_{22}+R_{33}+R_{44}.
\end{equation}
It follows that
\begin{equation}
   16\pi p_r(r)=R_{00}+R_{11}
  -R_{22}-R_{33}-R_{44}
\end{equation}
and, as can be readily checked,
\begin{equation}\label{E:tension2}
   16\pi p_r(r)=-\frac{1}{2r^2}\frac
   {\partial\mu(r,l)}{\partial r}
   \left[rb'(r)-b(r)\right]-
      \frac{2b(r)}{r^3}.
\end{equation}
We conclude that if $\partial\mu(r,l)
/\partial r<0$, Eq. (\ref{E:tension2})
is a tension whose magnitude depends on
$\partial\mu(r,l)/\partial r$.  In
principle, then, $\tau(r)$ can be
extremely large.

Our final task is to show that the model
can be made consistent with observation.
More precisely, we wish the extra dimension
to be extremely small or even curled up,
as in string theory.  To avoid a needless
complication, we would also like the extra
dimension to be essentially fixed from the
four-dimensional perspective.  For this
purpose, the form chosen, $e^{2\mu(r,l)}
dl^2$, is particularly convenient: assume
that $\mu(r,l)$ is negative and large in
absolute value.  This causes the proper
distance $e^{\mu(r,l)}dl$ and hence the
fifth component to be extremely small,
even negligible; the outcome is thereby
consistent with observation.  Now consider
\begin{equation}
   \frac{\partial}{\partial r}e^{\mu(r,l)}
   =e^{\mu(r,l)}\frac{\partial\mu(r,l)}
   {\partial r};
\end{equation}
we already know that $|\partial\mu(r,l)/
\partial r|$ can be extremely large.  At
the same time, $e^{\mu(r,l)}$ can be
extremely small; in fact, given the above
assumptions, $e^{\mu(r,l)}$ can be close to
Planck size.  Observationally, then,
\begin{equation}
   \frac{\partial}{\partial r}
   e^{\mu(r,l)}\approx 0.
\end{equation}
Since there is no discernable change in the
$r$-direction, the fifth dimension is
essentially fixed from the four-dimensional
perspective.

\section{Conclusion}
 This paper starts by recalling that
 a Morris-Thorne wormhole can only be
 held open by the use of exotic matter
 whose sole function is to ensure that
 the NEC is violated.  We have seen,
 however, that even a moderately-sized
 wormhole with $r_0=3\,\,\text{km}$ has
 a radial tension equal to that of the
 interior of a massive neutron star.
 Attributing this outcome to exotic
 matter merely gives the mystery a
 name since exotic matter was
 introduced for a completely
 different reason and with possibly
 different outcomes.  The purpose
 of this paper is to account for
 the large radial tension in three
 ways:  (1) directly appealing to
 noncommutative geometry, itself an
 offshoot of string theory, (2)
 invoking noncommutative geometry
 in conjunction with $f(R)$ modified
 gravity, and (3) determining the
 possible effect of a small extra
 spatial dimension.  All three models
 help confirm that whenever we are
 dealing with this extreme regime,
 the effects of string theory
 cannot be neglected \cite{lS93}.


\begin{thebibliography}{20}

\bibitem{MT88}M.S. Morris, K.S. Thorne,
   Am. J. Phys. \textbf{56}, 395 (1988).
\bibitem{pK12a}P.K.F. Kuhfittig, Adv. Math.
   Phys. \textbf{2013} 630196 (2012).
\bibitem{mV95}M. Visser, Lorentzian Wormholes:
   From Einstein to Hawking,\\ American
   Institute of Physics, New York, 1995) page 146.
\bibitem{eW96}E. Witten, Nucl. Phys. B
    \textbf{460} 335 (1996).
\bibitem{SW99}N. Seiberg, E. Witten,
   J. High Energy Phys. \textbf{9909}, 032 (1999).
\bibitem{NSS06}P. Nicolini, A. Smailagic, E.
    Spallucci, Phys. Lett. B \textbf{632}, 547 (2006).
\bibitem{NS10}P. Nicolini, E. Spallucci, Class.
   Quant. Grav. \textbf{27}, 015010 (2010).
\bibitem{mR11}M. Rinaldi, Class. Quant. Grav.
   \textbf{28}, 105022 (2011).
\bibitem{RKRIb}F. Rahaman, P.K.F. Kuhfittig, S. Ray,
   S. Islam, Phys. Rev. D \textbf{86}, 106101 (2012).
\bibitem{pK13}P.K.F. Kuhfittig, Int. J. Pure Appl.
   Math. \textbf{89}, 401 (2013).
\bibitem{LL12}J. Liang, B. Liu, Europhys. Lett.
   \textbf{100}, 30001 (2012).
\bibitem{NM08}K. Nozari, S.H. Mehdipour,
   Class. Quant. Grav. \textbf{25} 175015 (2008).
\bibitem{pK12}P.K.F. Kuhfittig, Adv. High Energy Phys.
   \textbf{2012}, 462493 (2012).
\bibitem{SS02}A. Smailagic, E. Spallucci, Phys. Rev. D
   \textbf{65}, 107701 (2002).
\bibitem{SS03a}A. Smailagic, E. Spallucci, J. Phys. A
   \textbf{36}, L-467 (2003).
\bibitem{SS03b}A. Smailagic, E. Spallucci, J. Phys. A
   \textbf{36}, L-517 (2003).
\bibitem{RJ16}S. Rani, A. Jawad, Adv. High Energy Phys.
   \textbf{2016}, 7815242 (2016).
\bibitem{pK15}P.K.F. Kuhfittig, Int. J. Mod. Phys. D
   \textbf{24}, 1550023 (2015).
\bibitem{LO09}F.S.N. Lobo, M.A. Oliveira,
   Phys. Rev. D \textbf{80} 104012 (2009).
\bibitem{pK18}P.K.F. Kuhfittig, New Horizons in
    Mathematical  Physics (NHMP) \textbf{2}, 62 (2018).
\bibitem{pK18a}P.K.F. Kuhfittig, Phys. Rev. D
   \textbf{98}, 064041 (2018).
\bibitem{lS93}L. Susskind, Phys. Rev. Lett.
   \textbf{71}, 2367 (1993).

 \end{thebibliography}
\end{document}